\documentclass[twocolumn,longbibliography,preprintnumbers,amsmath,amssymb,superscriptaddress]{revtex4-1}
\usepackage{graphicx}
\usepackage{bm}
\usepackage[dvipsnames]{xcolor}
\usepackage{color}
\usepackage{braket}
\usepackage{amsmath}
\usepackage{amssymb}
\usepackage{amsfonts}
\usepackage[english]{babel}
\usepackage{enumitem}
\usepackage[colorlinks]{hyperref}

\renewcommand{\d}{\mathrm d}

\begin{document}

\title{Nonequilibrium spin noise in a quantum dot ensemble}

\author{D.~S. Smirnov}
\affiliation{Ioffe Institute, 194021 St. Petersburg, Russia}

\author{Ph. Glasenapp}
\affiliation{Experimentelle Physik 2, Technische Universit\"at Dortmund, 44221 Dortmund, Germany}

\author{M. Bergen}
\affiliation{Experimentelle Physik 2, Technische Universit\"at Dortmund, 44221 Dortmund, Germany}

\author{M.~M. Glazov}
\affiliation{Ioffe Institute, 194021 St. Petersburg, Russia}

\author{D. Reuter}
\affiliation{Universit\"at Paderborn, Department Physik, 33098 Paderborn, Germany}

\author{A.~D. Wieck}
\affiliation{Angewandte Festk\"orperphysik, Ruhr-Universit\"at Bochum, 44780 Bochum, Germany}

\author{M. Bayer}
\affiliation{Experimentelle Physik 2, Technische Universit\"at Dortmund, 44221 Dortmund, Germany}
\affiliation{Ioffe Institute, 194021 St. Petersburg, Russia}

\author{A. Greilich}
\affiliation{Experimentelle Physik 2, Technische Universit\"at Dortmund, 44221 Dortmund, Germany}


\begin{abstract}
The spin noise in singly charged self-assembled quantum dots is studied theoretically and experimentally under the influence of a perturbation, provided by additional photoexcited charge carriers. The theoretical description takes into account generation and relaxation of charge carriers in the quantum dot system. The spin noise is measured under application of above barrier excitation for which the data are well reproduced by the developed model. Our analysis demonstrates a strong difference of the recharging dynamics for holes and electrons in quantum dots.
\end{abstract}


\maketitle

\label{sec:intro}
\textit{Introduction.}
Carrier spins localized in low-dimensional semiconductor nanostructures are currently one of the most promising objects for potential applications in quantum information devices~\cite{LloydScience93,LaddNature10}. In many respects, the interest in these systems and, in particular, in carrier spins in quantum dots (QDs) as discussed here, is associated with the observed long spin polarization lifetime and spin coherence time, caused by the relatively weak coupling of the spin degrees of freedom to lattice vibrations~\cite{QuantumBitsBook08}. At the same time, the large optical transition dipole moment and the interaction of orbital and spin degrees of freedom provide an excellent platform for coupling of spins to photons and make it possible to control the spin state by ultra-fast optical or electrical pulses~\cite{chapter6}.

Any practical realization based on such structures requires detailed information on the interaction of the spin system with its surrounding. It is highly desirable to get access to this information without any additional perturbation. As follows from the fluctuation-dissipation theorem, the dynamics of the system can be extracted from the spectrum of its fluctuations in thermodynamic equilibrium without any deliberate external perturbation, see Ref.~\cite{ChandlerBook87}. An experimental possibility for such a measurement is provided by the technique of spin noise spectroscopy~\cite{Alexandrov81,Zapasskii:13}.

Experimental studies on singly charged (In,Ga)As QDs using this technique have been performed by several groups~\cite{CrookerPRL2010,DahbashiAPL12}. In particular, it was shown, that the nuclear surrounding has a strong influence on the spectral response for both types of carriers, electrons and holes~\cite{HackmannPRL15}. The nuclear effects can be suppressed by a longitudinal external magnetic field, that stabilizes the carrier spin and decouples it from the nuclear bath~\cite{LiPRL12,GlasenappPRB16}. In all of these studies special care was taken to be as close as possible to the non-perturbative regime with the laser excitation. This is particularly important in optical studies of spin systems in semiconductors, for which even a tiny real excitation of the sample, accompanied by photo-injection of carriers, may drastically change the dynamics of the spin subsystem. As a result, in presence of excitation, the spin noise spectrum is, in general, no longer directly related with any spin response function~\cite{glazov:sns:jetp16, noise-excitons}. Hence, realization of a nonequilibrium spin noise situation deserves special attention and requires careful analysis.

In this paper we address theoretically and experimentally the question how the additional perturbation of the system by non-resonant excitation affects the  noise signal monitoring the ground state dynamics of carrier spins. We develop a minimal theoretical model, which describes the spin noise in non-equilibrium. The theoretical predictions are confirmed by the experiments. As a perturbation we use an above barrier excitation by a continuous wave laser, which injects additional carriers into the system.

\label{sec:theory}
\textit{Theory.}
The aim of the model is to describe the spin noise in an ensemble of QDs under non-resonant excitation. The QDs are singly charged with an electron or heavy hole. In both cases the ground state, $\left|g\right\rangle$, is two-fold degenerate with respect to the spin projection along the growth axis $z$. The non-resonant excitation can result in
(i) formation of a doubly-charged state if, e.g., one of the photocreated carriers is trapped outside the dot while the second one relaxes into the dot;
(ii) formation of an exciton;
(iii) formation of a charged exciton (trion). We neglect the contribution of the excited states to the noise signal. Indeed, for doubly-charged states the total spin is zero so that these states play no role at all. For the exciton state the optical resonance energy is quite different from the probe photon energy in  continuous wave measurements. For the sake of simplicity we also neglect the contribution of the trion states to the noise signal, assuming them weakly populated during the measurement time. As a result, only the resident carrier contributes to the spin noise, while the excited states can be treated as a reservoir.

\begin{figure}[t]
  \includegraphics[width=0.9\linewidth]{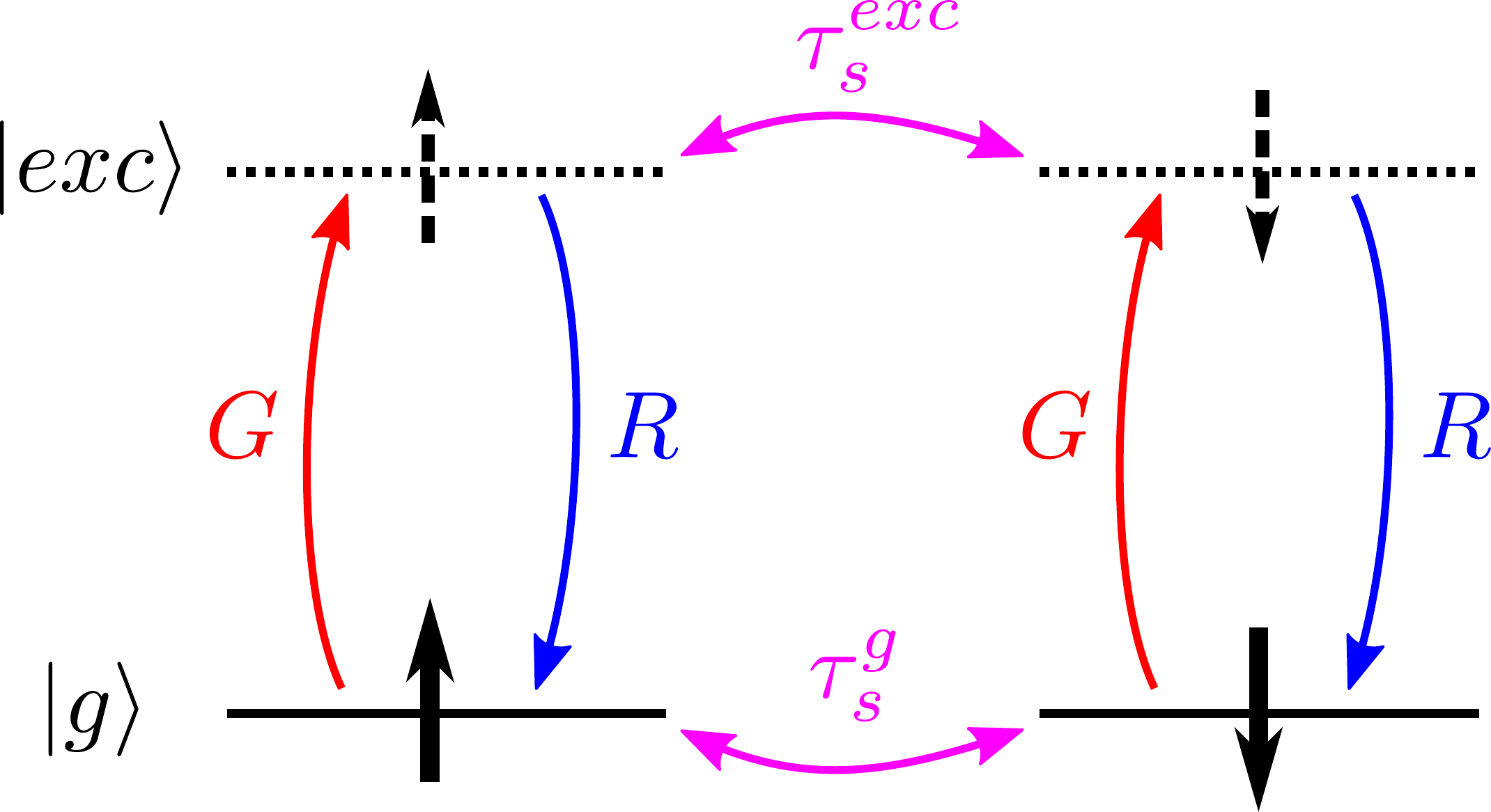}
\caption{Scheme of the transitions between ground, $|g\rangle$, and excited, $|exc\rangle$, states in the quantum dot and the spin relaxation processes in both states. The thick solid and dashed arrows show the spin projection, $S_z^{g,exc}$, in the ground and excited states, respectively.
}
\label{fig:levels}
\end{figure}

Optical pumping leads to excitation of the QD from the ground state, $|g\rangle$, to the reservoir of excited states, $|exc\rangle$. We present the minimal model and abstain from the microscopic description of carriers' photogeneration and diffusion to the quantum dots and of quantum dot recharging.
These processes are accounted for by two phenomenological parameters: the generation rate $G$ and the recombination (recharging) rate $R$, which describes the transitions from the excited back into the ground state, see Fig.~\ref{fig:levels}.
We assume that both rates, $G$ and $R$, are spin conserving and do not depend on the spin orientation of the resident charge carrier. The occupancy of the quantum dot ground state, $n_g$, obeys the kinetic equation
\begin{equation}
  \frac{\d n_g}{\d t}=-G n_g+Rn_{exc},
\end{equation}
where $n_{exc}=1-n_g$ is the occupancy of the excited states. In steady state one finds
\begin{equation}
  n_g=\frac{R}{R+G}.
  \label{eq:n}
\end{equation}
In what follows the recombination rate $R$ is kept constant and the generation rate $G$ is taken to be directly proportional to the pump power.

On average the spins of electrons and holes in the quantum dots are not polarized. The nonequilibrium spin noise is described by the matrix of correlation functions of the spin fluctuations in the ground, $\delta S_z^g$, and in the excited, $\delta S_z^{exc}$, states~\cite{glazov:sns:jetp16}
\begin{equation}
\label{M:dS}
\mathcal M(\tau) = \begin{pmatrix}
\langle \delta S_z^g(t+\tau) \delta S_z^g(t) \rangle & \langle \delta S_z^g(t+\tau) \delta S_z^{exc}(t) \rangle\\
\langle \delta S_z^{exc}(t+\tau) \delta S_z^g(t) \rangle & \langle \delta S_z^{exc}(t+\tau) \delta S_z^{exc}(t) \rangle
\end{pmatrix}
.
\end{equation}
Here the angular brackets denote averaging over $t$ at a fixed $\tau$. 

The electron and hole spin decoherence in quantum dots at cryogenic temperatures is usually governed by the hyperfine interaction with the host lattice nuclei. In $n$-type quantum dots, it results in a spin noise spectrum consisting of two peaks, one at the characteristic electron precession frequency in the field of nuclear spin fluctuations and one at $\omega=0$, related with the electron spins oriented parallel to the nuclear spin fluctuations. In $p$-type quantum dots, due to the hyperfine interaction anisotropy the spin noise spectrum is mainly concentrated at $\omega=0$~\cite{gi2012noise}. Here we focus on the zero-frequency peak in the spin noise spectrum, which can be easier accessed experimentally due to the limited detection range. This peak has a Lorentzian shape and can be described by a relaxation time, which is not related with the spin precession in the Overhauser field.

Bearing in mind these limitations, the matrix $\mathcal M(\tau)$ for $\tau>0$ obeys the kinetic equation
\begin{equation}
\label{M:kin}
\frac{\mathrm d}{\mathrm d\tau} \mathcal M(\tau) +
\begin{pmatrix}
G + 1/\tau_s^g & -R\\
-G & R + 1/\tau_s^{exc}
\end{pmatrix} \mathcal M(\tau) = 0,
\end{equation}
where $\tau_s^g$ and $\tau_s^{exc}$ are the spin relaxation times in the corresponding states, as sketched in Fig.~\ref{fig:levels}. Note that since the excited states are treated as a large reservoir, $\tau_s^{exc}$ is the effective phenomenological spin relaxation time in the excited states. For example, it can account for the recombination processes without spin conservation.

The initial conditions for the matrix $\mathcal M(\tau)$ follow from the single time correlators
\begin{equation}
\label{eq:sz2}
\left\langle (\delta S_z^g)^2\right\rangle=\frac{n_g}{4}, \quad \left\langle (\delta S_z^{exc})^2\right\rangle=\frac{n_{exc}}{4},
\end{equation}
and $\langle \delta S_z^g \delta S_z^{exc}\rangle=0$. Thus, Eq.~\eqref{M:kin} can be readily solved, and the spin noise spectrum $\left(\delta S_z^2\right)_\omega$ can be found in the standard way as the Fourier transform of the correlator $\langle \delta S_z^g(t+\tau) \delta S_z^g(t) \rangle$
~\cite{glazov:sns:jetp16,noise-excitons,SinitsynReview}.

Provided the recombination rate is faster than the spin relaxation rates, $R^{-1}\ll\tau_s^{g,exc}$, we arrive at the spin noise spectrum in the ground state of a single QD,~cf. Refs.~\cite{glazov:sns:jetp16,PhysRevB.89.081304}:
\begin{equation}
\label{spectrum}
  \left(\delta S_z^2\right)_\omega=\frac{n_g^2}{2}\frac{\tau_s^*}{1+(\omega\tau_s^*)^2}
  +\frac{n_gn_{exc}}{2}\frac{R+G}{(R+G)^2+\omega^2}
,
\end{equation}
where we have introduced an average spin relaxation time $\tau_s^*$ according to
\begin{equation}
  \frac{1}{\tau_s^*}=\frac{n_g}{\tau_s^g}+\frac{n_{exc}}{\tau_s^{exc}}.
\label{eq:w}
\end{equation}
The spectrum consists of a pair of Lorentzian peaks centered at zero frequency. The peak, described by the second term of Eq.~\eqref{spectrum} is related to the fast generation-recombination processes. It is much wider than the other one, and expected to be beyond the frequency range in the measurements. Thus the area of the spectrum $\propto n_g^2$ and with an increase of the pumping rate $G$ decreases faster than the occupancy of the ground state. The width of the spin noise spectrum is the weighted average of the spin decay rates in the ground and the excited states. In the following, we test the theoretical predictions experimentally.

\label{sec:SaS}
\textit{Setup and samples.}
The experimental setup and samples are presented in detail in the Ref.~\cite{GlasenappPRB16}. Here we shortly recapitulate important characteristics of the samples and the setup and complete the description. We study QD ensembles, where the resident carriers are represented either by holes or by electrons. Both samples contain 20 layers of MBE-grown (In,Ga)As/GaAs QDs separated by 60\,nm GaAs barriers. The QD density is $n=10^{10}$~cm$^{-2}$ per layer. The $p$-doped sample was not intentionally doped, but has a background level of $p$-type doping due to residual carbon impurities in the growth chamber. The $n$-doping was obtained by incorporating $\delta$-sheets of Si dopants 20\,nm below each QD layer, with the dopant density roughly equal to the QD density. The samples are mounted on the cold finger of a liquid Helium flow cryostat, and cooled down to a temperature of 5\,K.

In order to detect the spin fluctuations we measure the noise of the polarization plane rotation of the probe beam transmitted through the sample. The linearly polarized probe light is taken from a tunable continuous wave (\textit{cw}) Titanium-Sapphire ring laser emitting a single frequency laser line with a linewidth  <\,10\,MHz. The laser is then guided through a single mode optical fiber for spatial mode shaping ($\text{TEM}_{00}$). After a Glan-Taylor polarizer providing a linear extinction ratio of 10$^5$, the beam is transmitted through the sample. The probe laser power density at the sample was fixed at a level of 45\,W/cm$^2$ (1\,mW with 53\,$\mu$m spot diameter), which was selected as a compromise between signal level above shot noise versus perturbation of the system caused by the laser. For the $n$-doped sample we use the probe wavelength of 889\,nm and for the $p$-doped sample of 895\,nm within the ground state absorption bands, which correspond to the spectral positions of maximum spin noise signals, see Ref.~\cite{GlasenappPRB16}.

Detection of the Faraday rotation noise is done with a standard polarimeter, consisting of a Wollaston prism and a broadband optical balanced detector with the bandwidth of 100\,MHz (Femto HCA-S)~\cite{GlasenappPRB13}. The output of the detector is amplified and low-pass filtered at 60\,MHz to avoid undersampling, and then sent to the input of a digitizer with the incorporated field programmable gate array for real time Fast Fourier Transform computation.

In order to remove the spin-independent contribution from the overall photon shot noise and the intrinsic electronic noise, the spin noise measurements were interleaved between two magnetic fields: zero field and the $B_{x} = 100$\,mT in Voigt geometry for shifting the Larmor precession peak out of the detectable bandwidth. Subtraction of both spectra then yields the pure spin noise contribution. To account for the frequency dependent sensitivity of the diodes we normalize the spectra by the photon shot noise power spectra.

Further, we provide above barrier excitation by a \textit{cw} diode laser operating at the wavelength of 785\,nm (pump laser). The laser is linearly polarized and directed to the sample at small angle relative to the probe beam. At this wavelength it is completely absorbed in the bulk of the sample, so that no pump light exits the sample. The laser is not focused by any lens and its spot size is measured to be 1.2\,mm. So the pump spot at the sample is much larger than the probe spot, which allows us to have good homogeneity of the excitation in the probed area.

\begin{figure}[t]
\includegraphics[width=\columnwidth]{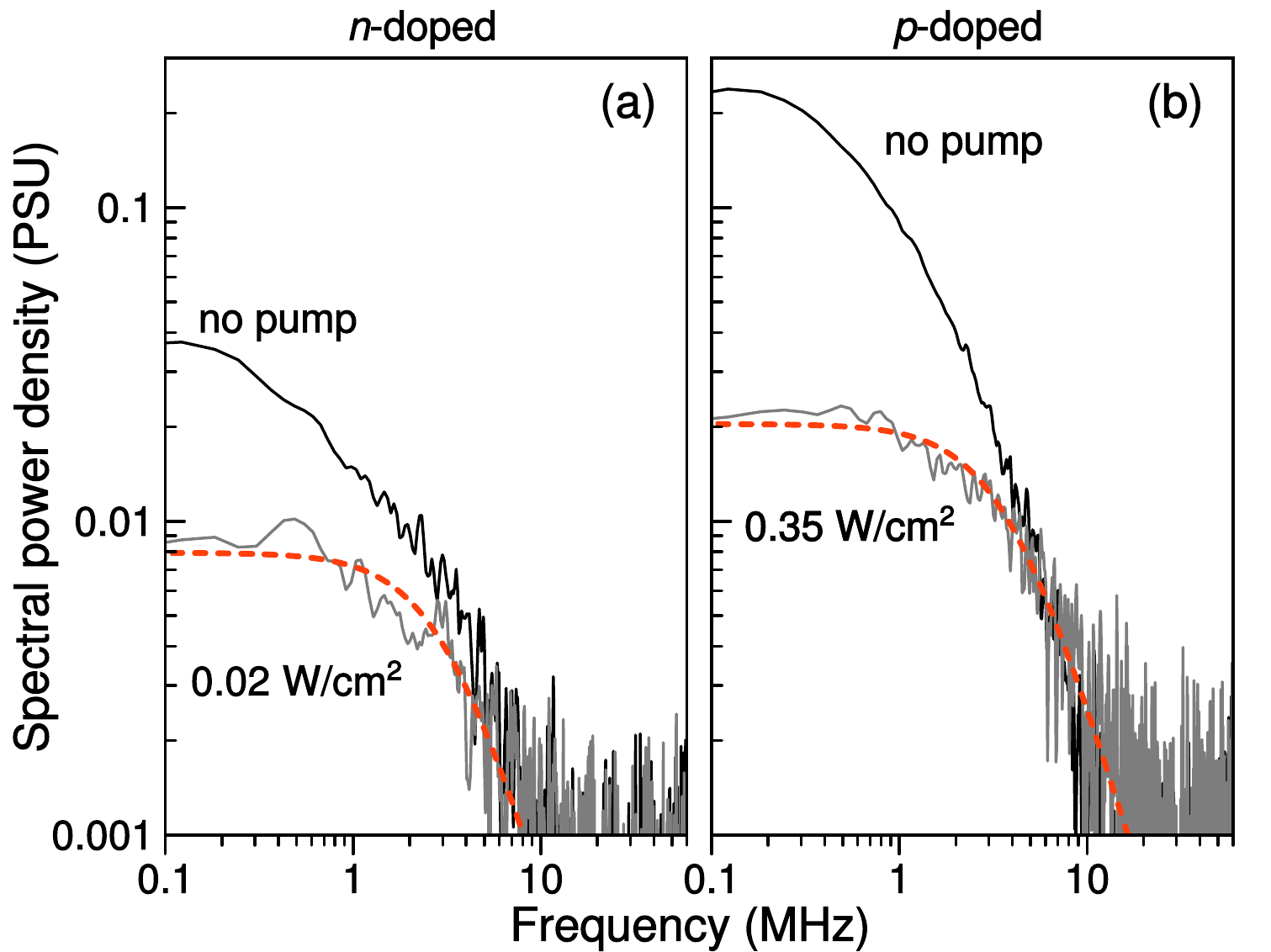}
\caption{Spin noise spectra in units of photon shot noise (PSU) with and without additional above barrier excitation, as indicated in the figure, measured in (a) $n$-doped and (b)~$p$-doped sample. The red dashed lines show Lorentzian fits.
}\label{fig:spectra}
\end{figure}

\label{sec:Exp}
\textit{Experimental results.}
Figure~\ref{fig:spectra} presents measured spectra for the $n$- and $p$-doped samples at $B=0$ for zero and nonzero
pump power. The observed phenomenology is overall the same for both types of samples: The spectral noise area decreases and the half width at half maximum (HWHM) of the peaks increases with increasing power density. The frequency dependence of all spectra closely follows the Lorentzian form, as demonstrated by exemplary fits shown by the red dashed lines. This indicates that the features in the spin noise spectra that are related to the spin precession in the Overhauser field are beyond the detection range, in agreement with the model assumptions.

\begin{figure}[t]
\includegraphics[width=\columnwidth]{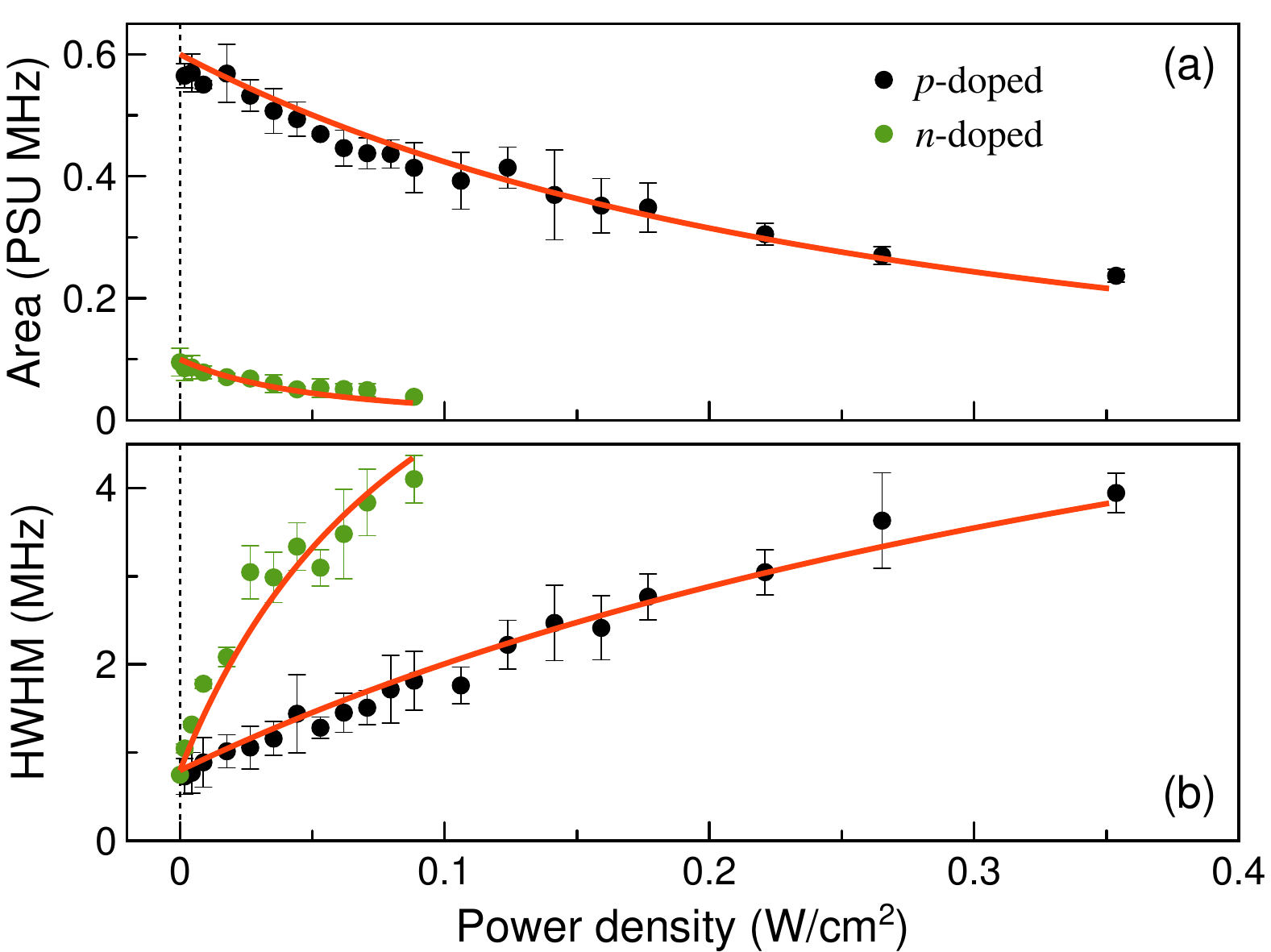}
\caption{(a) The area under the SNS in photon shot noise units as function of pump density for $n$-doped (green points) and $p$-doped (black points) samples. (b) Half width at half maximum of the spin noise spectrum. The measurements are repeated 3 times at each pump power, which leads to the presented error bars. The red solid lines are fits by Eq.~\eqref{spectrum}, see the main text for the details.}\label{fig:fits}
\end{figure}

The spin noise signal in all cases decays within 60\,MHz range down to the shot noise level, so that we limit our fit to this range. From the Lorentzian fit we extract the characteristic parameters of the peak: the full spectral area and the HWHM. These data are presented in the Fig.~\ref{fig:fits}. The areas are very different at zero pump power for the electron and hole doped sample which is related to two factors: (i) most of the spin noise power for the $n$-doped sample is concentrated outside the zero frequency peak and (ii) the concentration of charge carriers and the optical transition dipole moment are somewhat larger for the $p$-doped sample~\cite{GlasenappPRB16}.

The power dependence of the area for both carrier types could be related to the capture efficiency of additional carriers into the QDs. For example, let us  consider the $n$-doped QD: Due to a capture process a singly charged QD becomes either populated by two electrons forming a singlet state, which does not produce noise signal, or by an exciton, once a hole is captured. If the electron-hole complex forms a bright exciton, it has a recombination time of about 400\,ps~\cite{GreilichPRB73}, which is outside of the detection bandwidth of the current setup. If the exciton is in the dark state, the recombination times can reach several microseconds and could in principle contribute to the signal if the times do not exceed 10 microseconds, the upper limit of our setup. However, the exciton resonance frequency for the considered quantum dot is significantly different from the trion frequency, so that in any case the formation of excitons in the QD effectively switches off its contribution to this noise, in agreement with the theoretical model. In (In,Ga)As QDs the energy difference between the exciton and trion transitions is usually at the level of several meV~\cite{PhysRevLett.79.5282}, while the homogeneous linewidth, which is relevant for the spin noise, is at the order of several $\mu$eV~\cite{PhysRevLett.110.176601,PhysRevB.90.205306,NatComm14} at temperatures of a few K. Below we show how a difference in the HWHM can be related to differences in the carrier generation and/or relaxation kinetics. 

\label{sec:discus}
\textit{Discussion.}
Figure~\ref{fig:fits} shows fits to the areas and the widths of the spin noise spectra for the two studied samples after the theoretical model presented above. The area is assumed to be proportional to $n_g^2$, see Eqs.~\eqref{eq:n} and~\eqref{spectrum}, and the HWHM is given by Eq.~\eqref{eq:w}. Note that the generation and recombination rates cannot be separately determined from the fit, but only their ratio, $G/R$, which describes the spin noise spectrum, can be extracted. As noted above, we assume the simplest model, where the recombination rate does not depend on the pump power, $P$, while the generation rate is linearly proportional to it. More complicated dependencies can be obtained by taking into account multiple excited states and transitions between them, which is beyond the minimal model presented
here~\footnote{Assuming that the pump beam is completely absorbed, and all the photoexcited carriers are captured in the topmost QD layer we estimate $G\approx P/(\hbar\omega n)$, where $\omega$ is the frequency of the pump light. For the $p$-doped sample the generation and recombination rates become comparable at $P=0.5$~W/cm$^2$. This value corresponds to $R\approx G=0.2$~ns$^{-1}$, which supports our assumption that the recombination and generation processes at high pump power are faster than the localized carrier spin relaxation. For the $n$-doped sample, however, the estimate gives comparable values of $R^{-1}$ and $\tau_s^{exc}$. Our aim is to describe the spin noise spectra rather than the generation-recombination dynamics, so we use Eq.~\eqref{spectrum} to fit both the electron and hole spin noise spectra.}.

From the fits we have determined the spin relaxation times in the ground and excited states, as well as the constant coefficient $G/(RP)$. The results are summarized in Tab.~\ref{tab:params}. For both samples the spin relaxation times in the excited states are shorter than in the ground states, as expected. Therefore excitation of the QDs leads to shortening of the average spin relaxation time and broadening of the spin noise spectra.

The spin relaxation times in the ground states are somewhat shorter than the values measured in Ref.~\cite{GlasenappPRB16} for the same samples. This is related to the higher probe power density (45\,W/cm$^2$ vs. 1\,W/cm$^2$) as compared to Ref.~\cite{GlasenappPRB16}. Interestingly the spin relaxation times for the ground and excited states, independently determined for the two samples, are almost identical within the experimental accuracy. 
The spin relaxation in the ground state in the absence of additional excitation increases with increasing probe power faster for the $n$-doped sample than for the $p$-doped~\cite{GlasenappPRB16}. As the probe powers used here are quite large, we conclude that the spin relaxation time in the limit of small probe power would be longer for the $n$-doped sample.

The coincidence of spin relaxation times in the excited states suggests that the loss of spin polarization is related to charge kinetics, e.g., to escape from the dot or recombination with another carrier. The dynamics of these processes can be similar for the two samples under study. A detailed analysis of these processes is beyond our model.

 \begin{table}[t]
 \vspace{-0.3cm} \caption{The parameters of spin and charge dynamics for the $n$- and $p$-doped samples determined from the fits.}
 \label{tab:params}
 \begin{ruledtabular}
 \begin{tabular}{ccccccc}
 & $n$-doped & $p$-doped \\
 \hline
 $\tau_s^g$\,(ns) & 200 & 200 \\
 $\tau_s^{exc}$\,(ns) & 19 & 19 \\
 $G/(RP)$\,(cm$^2$/W) & 10 & 1.9 \\
 \end{tabular}
 \end{ruledtabular}
 \end{table}

The main difference between the two samples is the faster increase of the spin relaxation rate with increasing pump power for the $n$-doped sample than for the $p$-doped sample, as clearly seeing in Fig.~\ref{fig:fits}(b). In detail, the ratio of generation and recombination rates determined from the fits is roughly five times larger. The generation is related to the excitation of additional carriers and their capture in the QDs, so that we expect the generation rate to be similar for the two samples. Hence we conclude, that the recombination rate for the $p$-doped sample is larger. This can be related to the smaller spacings between the size-quantized levels for holes than for electrons~\cite{Kurtze2009}.

We note that the nominally $n$-doped sample contains a fraction of positively charged QDs~\cite{GlasenappPRB16}. Provided that the spin relaxation for holes is faster than for electrons, the HWHM can be determined by the corresponding shortest time even for a small fraction of positively charged QDs. An increase of the pump power accelerates the spin relaxation which does not change the situation. Therefore the similar spin relaxation times in the ground and excited states for $p$- and $n$-doped samples can be alternatively explained by the small concentration of holes in the nominally $n$-doped sample.

\label{sec:conclusion}
\textit{Conclusion.}
We have studied the spin noise in ensembles of charged QDs in strong nonequilibrium conditions, when nonresonant excitation provides additional carriers. The proposed minimal four level model, which takes into account the excitation and recombination dynamics, allows for the analytical description of the spin noise spectra. We have found that the excitation leads to a decrease of the spin noise power and a broadening of the spectrum.
The theoretical predictions are supported by the experimental measurements of $n$- and $p$-type QDs in presence of above barrier excitation. From the fit of experimental data the parameters of spin relaxation in the ground and excited states have been determined and the excitation-recombination dynamics have been analyzed. Qualitatively, the behavior of the $p$- and $n$-type samples is the same, the main quantitative difference  is dictated by the difference in the recharging rates of the dots.

\textit{Acknowledgments.}
We acknowledge the financial support by the Deutsche Forschungsgemeinschaft in the frame of the ICRC TRR 160 and RFBR-DFG grant 15-52-12012, RFBR grant 16-32-00540, RF President Grants No. SP-643.2015.5 and MD-1555.2017.2, and Federation Government Grant No. 14.Z50.31.0021 (leading scientist M. Bayer).


\begin{thebibliography}{25}%
\makeatletter
\providecommand \@ifxundefined [1]{%
 \@ifx{#1\undefined}
}%
\providecommand \@ifnum [1]{%
 \ifnum #1\expandafter \@firstoftwo
 \else \expandafter \@secondoftwo
 \fi
}%
\providecommand \@ifx [1]{%
 \ifx #1\expandafter \@firstoftwo
 \else \expandafter \@secondoftwo
 \fi
}%
\providecommand \natexlab [1]{#1}%
\providecommand \enquote  [1]{``#1''}%
\providecommand \bibnamefont  [1]{#1}%
\providecommand \bibfnamefont [1]{#1}%
\providecommand \citenamefont [1]{#1}%
\providecommand \href@noop [0]{\@secondoftwo}%
\providecommand \href [0]{\begingroup \@sanitize@url \@href}%
\providecommand \@href[1]{\@@startlink{#1}\@@href}%
\providecommand \@@href[1]{\endgroup#1\@@endlink}%
\providecommand \@sanitize@url [0]{\catcode `\\12\catcode `\$12\catcode
  `\&12\catcode `\#12\catcode `\^12\catcode `\_12\catcode `\%12\relax}%
\providecommand \@@startlink[1]{}%
\providecommand \@@endlink[0]{}%
\providecommand \url  [0]{\begingroup\@sanitize@url \@url }%
\providecommand \@url [1]{\endgroup\@href {#1}{\urlprefix }}%
\providecommand \urlprefix  [0]{URL }%
\providecommand \Eprint [0]{\href }%
\providecommand \doibase [0]{http://dx.doi.org/}%
\providecommand \selectlanguage [0]{\@gobble}%
\providecommand \bibinfo  [0]{\@secondoftwo}%
\providecommand \bibfield  [0]{\@secondoftwo}%
\providecommand \translation [1]{[#1]}%
\providecommand \BibitemOpen [0]{}%
\providecommand \bibitemStop [0]{}%
\providecommand \bibitemNoStop [0]{.\EOS\space}%
\providecommand \EOS [0]{\spacefactor3000\relax}%
\providecommand \BibitemShut  [1]{\csname bibitem#1\endcsname}%
\let\auto@bib@innerbib\@empty
\bibitem [{\citenamefont {Lloyd}(1993)}]{LloydScience93}%
  \BibitemOpen
  \bibfield  {author} {\bibinfo {author} {\bibfnamefont {Seth}\ \bibnamefont
  {Lloyd}},\ }\bibfield  {title} {\enquote {\bibinfo {title} {A potential
  realizable quantum computer},}\ }\href@noop {} {\bibfield  {journal}
  {\bibinfo  {journal} {Science}\ }\textbf {\bibinfo {volume} {261}},\ \bibinfo
  {pages} {1569} (\bibinfo {year} {1993})}\BibitemShut {NoStop}%
\bibitem [{\citenamefont {Ladd}\ \emph {et~al.}(2010)\citenamefont {Ladd},
  \citenamefont {Jelezko}, \citenamefont {Laflamme}, \citenamefont {Nakamura},
  \citenamefont {Monroe},\ and\ \citenamefont {O'Brien}}]{LaddNature10}%
  \BibitemOpen
  \bibfield  {author} {\bibinfo {author} {\bibfnamefont {T.~D.}\ \bibnamefont
  {Ladd}}, \bibinfo {author} {\bibfnamefont {F.}~\bibnamefont {Jelezko}},
  \bibinfo {author} {\bibfnamefont {R.}~\bibnamefont {Laflamme}}, \bibinfo
  {author} {\bibfnamefont {Y.}~\bibnamefont {Nakamura}}, \bibinfo {author}
  {\bibfnamefont {C.}~\bibnamefont {Monroe}}, \ and\ \bibinfo {author}
  {\bibfnamefont {J.~L.}\ \bibnamefont {O'Brien}},\ }\bibfield  {title}
  {\enquote {\bibinfo {title} {Quantum computers},}\ }\href@noop {} {\bibfield
  {journal} {\bibinfo  {journal} {Nature}\ }\textbf {\bibinfo {volume} {464}},\
  \bibinfo {pages} {45} (\bibinfo {year} {2010})}\BibitemShut {NoStop}%
\bibitem [{\citenamefont {Henneberger}\ and\ \citenamefont
  {Benson}(2008)}]{QuantumBitsBook08}%
  \BibitemOpen
  \bibinfo {editor} {\bibfnamefont {Fritz}\ \bibnamefont {Henneberger}}\ and\
  \bibinfo {editor} {\bibfnamefont {Oliver}\ \bibnamefont {Benson}},\ eds.,\
  \href@noop {} {\emph {\bibinfo {title} {Semiconductor Quantum Bits}}}\
  (\bibinfo  {publisher} {Pan Stanford},\ \bibinfo {year} {2008})\BibitemShut
  {NoStop}%
\bibitem [{\citenamefont {Greilich}\ \emph {et~al.}(2010)\citenamefont
  {Greilich}, \citenamefont {Yakovlev},\ and\ \citenamefont
  {Bayer}}]{chapter6}%
  \BibitemOpen
  \bibfield  {author} {\bibinfo {author} {\bibfnamefont {A.}~\bibnamefont
  {Greilich}}, \bibinfo {author} {\bibfnamefont {D.~R.}\ \bibnamefont
  {Yakovlev}}, \ and\ \bibinfo {author} {\bibfnamefont {M.}~\bibnamefont
  {Bayer}},\ }\href@noop {} {\emph {\bibinfo {title} {Optical Generation and
  Control of Quantum Coherence in Semiconductor Nanostructures}}},\ edited by\
  \bibinfo {editor} {\bibfnamefont {G.}~\bibnamefont {Slavcheva}}\ and\
  \bibinfo {editor} {\bibfnamefont {Ph.}\ \bibnamefont {Roussignol}}\ (\bibinfo
   {publisher} {Springer-Verlag},\ \bibinfo {address} {Berlin},\ \bibinfo
  {year} {2010})\ Chap.~\bibinfo {chapter} {6}, p.~\bibinfo {pages}
  {85}\BibitemShut {NoStop}%
\bibitem [{\citenamefont {Chandler}(1987)}]{ChandlerBook87}%
  \BibitemOpen
  \bibinfo {editor} {\bibfnamefont {David}\ \bibnamefont {Chandler}},\ ed.,\
  \enquote {\bibinfo {title} {Introduction to modern statistical mechanics},}\
  \ (\bibinfo  {publisher} {Oxford University Press},\ \bibinfo {year} {1987})\
  Chap.\ \bibinfo {chapter} {Statistical Mechanics of Non-Equilibrium Systems},
  p.\ \bibinfo {pages} {255}\BibitemShut {NoStop}%
\bibitem [{\citenamefont {Aleksandrov}\ and\ \citenamefont
  {Zapasskii}(1981)}]{Alexandrov81}%
  \BibitemOpen
  \bibfield  {author} {\bibinfo {author} {\bibfnamefont {E.~B.}\ \bibnamefont
  {Aleksandrov}}\ and\ \bibinfo {author} {\bibfnamefont {V.~S.}\ \bibnamefont
  {Zapasskii}},\ }\bibfield  {title} {\enquote {\bibinfo {title} {Magnetic
  resonance in the faraday-rotation noise spectrum},}\ }\href@noop {}
  {\bibfield  {journal} {\bibinfo  {journal} {JETP 54, 64 (1981)}\ }\textbf
  {\bibinfo {volume} {54}},\ \bibinfo {pages} {64} (\bibinfo {year}
  {1981})}\BibitemShut {NoStop}%
\bibitem [{\citenamefont {Zapasskii}(2013)}]{Zapasskii:13}%
  \BibitemOpen
  \bibfield  {author} {\bibinfo {author} {\bibfnamefont {V.~S.}\ \bibnamefont
  {Zapasskii}},\ }\bibfield  {title} {\enquote {\bibinfo {title} {Spin-noise
  spectroscopy: from proof of principle to applications},}\ }\href {\doibase
  10.1364/AOP.5.000131} {\bibfield  {journal} {\bibinfo  {journal} {Adv. Opt.
  Photon.}\ }\textbf {\bibinfo {volume} {5}},\ \bibinfo {pages} {131} (\bibinfo
  {year} {2013})}\BibitemShut {NoStop}%
\bibitem [{\citenamefont {Crooker}\ \emph {et~al.}(2010)\citenamefont
  {Crooker}, \citenamefont {Brandt}, \citenamefont {Sandfort}, \citenamefont
  {Greilich}, \citenamefont {Yakovlev}, \citenamefont {Reuter}, \citenamefont
  {Wieck},\ and\ \citenamefont {Bayer}}]{CrookerPRL2010}%
  \BibitemOpen
  \bibfield  {author} {\bibinfo {author} {\bibfnamefont {S.~A.}\ \bibnamefont
  {Crooker}}, \bibinfo {author} {\bibfnamefont {J.}~\bibnamefont {Brandt}},
  \bibinfo {author} {\bibfnamefont {C.}~\bibnamefont {Sandfort}}, \bibinfo
  {author} {\bibfnamefont {A.}~\bibnamefont {Greilich}}, \bibinfo {author}
  {\bibfnamefont {D.~R.}\ \bibnamefont {Yakovlev}}, \bibinfo {author}
  {\bibfnamefont {D.}~\bibnamefont {Reuter}}, \bibinfo {author} {\bibfnamefont
  {A.~D.}\ \bibnamefont {Wieck}}, \ and\ \bibinfo {author} {\bibfnamefont
  {M.}~\bibnamefont {Bayer}},\ }\bibfield  {title} {\enquote {\bibinfo {title}
  {Spin noise of electrons and holes in self-assembled quantum dots},}\ }\href
  {\doibase 10.1103/PhysRevLett.104.036601} {\bibfield  {journal} {\bibinfo
  {journal} {Phys. Rev. Lett.}\ }\textbf {\bibinfo {volume} {104}},\ \bibinfo
  {pages} {036601} (\bibinfo {year} {2010})}\BibitemShut {NoStop}%
\bibitem [{\citenamefont {Dahbashi}\ \emph {et~al.}(2012)\citenamefont
  {Dahbashi}, \citenamefont {H\"ubner}, \citenamefont {Berski}, \citenamefont
  {Wiegand}, \citenamefont {Marie}, \citenamefont {Pierz}, \citenamefont
  {Schumacher},\ and\ \citenamefont {Oestreich}}]{DahbashiAPL12}%
  \BibitemOpen
  \bibfield  {author} {\bibinfo {author} {\bibfnamefont {R.}~\bibnamefont
  {Dahbashi}}, \bibinfo {author} {\bibfnamefont {J.}~\bibnamefont {H\"ubner}},
  \bibinfo {author} {\bibfnamefont {F.}~\bibnamefont {Berski}}, \bibinfo
  {author} {\bibfnamefont {J.}~\bibnamefont {Wiegand}}, \bibinfo {author}
  {\bibfnamefont {X.}~\bibnamefont {Marie}}, \bibinfo {author} {\bibfnamefont
  {K.}~\bibnamefont {Pierz}}, \bibinfo {author} {\bibfnamefont {H.~W.}\
  \bibnamefont {Schumacher}}, \ and\ \bibinfo {author} {\bibfnamefont
  {M.}~\bibnamefont {Oestreich}},\ }\bibfield  {title} {\enquote {\bibinfo
  {title} {Measurement of heavy-hole spin dephasing in $\rm{(In,Ga)As}$ quantum
  dots},}\ }\href@noop {} {\bibfield  {journal} {\bibinfo  {journal} {Appl.
  Phys. Lett.}\ }\textbf {\bibinfo {volume} {100}} (\bibinfo {year}
  {2012})}\BibitemShut {NoStop}%
\bibitem [{\citenamefont {Hackmann}\ \emph {et~al.}(2015)\citenamefont
  {Hackmann}, \citenamefont {Glasenapp}, \citenamefont {Greilich},
  \citenamefont {Bayer},\ and\ \citenamefont {Anders}}]{HackmannPRL15}%
  \BibitemOpen
  \bibfield  {author} {\bibinfo {author} {\bibfnamefont {J.}~\bibnamefont
  {Hackmann}}, \bibinfo {author} {\bibfnamefont {Ph.}\ \bibnamefont
  {Glasenapp}}, \bibinfo {author} {\bibfnamefont {A.}~\bibnamefont {Greilich}},
  \bibinfo {author} {\bibfnamefont {M.}~\bibnamefont {Bayer}}, \ and\ \bibinfo
  {author} {\bibfnamefont {F.~B.}\ \bibnamefont {Anders}},\ }\bibfield  {title}
  {\enquote {\bibinfo {title} {Influence of the nuclear electric quadrupolar
  interaction on the coherence time of hole and electron spins confined in
  semiconductor quantum dots},}\ }\href {\doibase
  10.1103/PhysRevLett.115.207401} {\bibfield  {journal} {\bibinfo  {journal}
  {Phys. Rev. Lett.}\ }\textbf {\bibinfo {volume} {115}},\ \bibinfo {pages}
  {207401} (\bibinfo {year} {2015})}\BibitemShut {NoStop}%
\bibitem [{\citenamefont {Li}\ \emph {et~al.}(2012)\citenamefont {Li},
  \citenamefont {Sinitsyn}, \citenamefont {Smith}, \citenamefont {Reuter},
  \citenamefont {Wieck}, \citenamefont {Yakovlev}, \citenamefont {Bayer},\ and\
  \citenamefont {Crooker}}]{LiPRL12}%
  \BibitemOpen
  \bibfield  {author} {\bibinfo {author} {\bibfnamefont {Yan}\ \bibnamefont
  {Li}}, \bibinfo {author} {\bibfnamefont {N.}~\bibnamefont {Sinitsyn}},
  \bibinfo {author} {\bibfnamefont {D.~L.}\ \bibnamefont {Smith}}, \bibinfo
  {author} {\bibfnamefont {D.}~\bibnamefont {Reuter}}, \bibinfo {author}
  {\bibfnamefont {A.~D.}\ \bibnamefont {Wieck}}, \bibinfo {author}
  {\bibfnamefont {D.~R.}\ \bibnamefont {Yakovlev}}, \bibinfo {author}
  {\bibfnamefont {M.}~\bibnamefont {Bayer}}, \ and\ \bibinfo {author}
  {\bibfnamefont {S.~A.}\ \bibnamefont {Crooker}},\ }\bibfield  {title}
  {\enquote {\bibinfo {title} {Intrinsic spin fluctuations reveal the dynamical
  response function of holes coupled to nuclear spin baths in (in,ga)as quantum
  dots},}\ }\href {\doibase 10.1103/PhysRevLett.108.186603} {\bibfield
  {journal} {\bibinfo  {journal} {Phys. Rev. Lett.}\ }\textbf {\bibinfo
  {volume} {108}},\ \bibinfo {pages} {186603} (\bibinfo {year}
  {2012})}\BibitemShut {NoStop}%
\bibitem [{\citenamefont {Glasenapp}\ \emph {et~al.}(2016)\citenamefont
  {Glasenapp}, \citenamefont {Smirnov}, \citenamefont {Greilich}, \citenamefont
  {Hackmann}, \citenamefont {Glazov}, \citenamefont {Anders},\ and\
  \citenamefont {Bayer}}]{GlasenappPRB16}%
  \BibitemOpen
  \bibfield  {author} {\bibinfo {author} {\bibfnamefont {Ph.}\ \bibnamefont
  {Glasenapp}}, \bibinfo {author} {\bibfnamefont {D.~S.}\ \bibnamefont
  {Smirnov}}, \bibinfo {author} {\bibfnamefont {A.}~\bibnamefont {Greilich}},
  \bibinfo {author} {\bibfnamefont {J.}~\bibnamefont {Hackmann}}, \bibinfo
  {author} {\bibfnamefont {M.~M.}\ \bibnamefont {Glazov}}, \bibinfo {author}
  {\bibfnamefont {F.~B.}\ \bibnamefont {Anders}}, \ and\ \bibinfo {author}
  {\bibfnamefont {M.}~\bibnamefont {Bayer}},\ }\bibfield  {title} {\enquote
  {\bibinfo {title} {Spin noise of electrons and holes in $\rm{(In,Ga)As}$
  quantum dots: Experiment and theory},}\ }\href {\doibase
  10.1103/PhysRevB.93.205429} {\bibfield  {journal} {\bibinfo  {journal} {Phys.
  Rev. B}\ }\textbf {\bibinfo {volume} {93}},\ \bibinfo {pages} {205429}
  (\bibinfo {year} {2016})}\BibitemShut {NoStop}%
\bibitem [{\citenamefont {Glazov}(2016)}]{glazov:sns:jetp16}%
  \BibitemOpen
  \bibfield  {author} {\bibinfo {author} {\bibfnamefont {M.~M.}\ \bibnamefont
  {Glazov}},\ }\bibfield  {title} {\enquote {\bibinfo {title} {Spin
  fluctuations of nonequilibrium electrons and excitons in semiconductors},}\
  }\href {\doibase 10.1134/S1063776116030067} {\bibfield  {journal} {\bibinfo
  {journal} {JETP}\ }\textbf {\bibinfo {volume} {122}},\ \bibinfo {pages} {472}
  (\bibinfo {year} {2016})}\BibitemShut {NoStop}%
\bibitem [{\citenamefont {Smirnov}\ and\ \citenamefont
  {Glazov}(2014)}]{noise-excitons}%
  \BibitemOpen
  \bibfield  {author} {\bibinfo {author} {\bibfnamefont {D.~S.}\ \bibnamefont
  {Smirnov}}\ and\ \bibinfo {author} {\bibfnamefont {M.~M.}\ \bibnamefont
  {Glazov}},\ }\bibfield  {title} {\enquote {\bibinfo {title} {Exciton spin
  noise in quantum wells},}\ }\href {\doibase 10.1103/PhysRevB.90.085303}
  {\bibfield  {journal} {\bibinfo  {journal} {Phys. Rev. B}\ }\textbf {\bibinfo
  {volume} {90}},\ \bibinfo {pages} {085303} (\bibinfo {year}
  {2014})}\BibitemShut {NoStop}%
\bibitem [{\citenamefont {Glazov}\ and\ \citenamefont
  {Ivchenko}(2012)}]{gi2012noise}%
  \BibitemOpen
  \bibfield  {author} {\bibinfo {author} {\bibfnamefont {M.~M.}\ \bibnamefont
  {Glazov}}\ and\ \bibinfo {author} {\bibfnamefont {E.~L.}\ \bibnamefont
  {Ivchenko}},\ }\bibfield  {title} {\enquote {\bibinfo {title} {Spin noise in
  quantum dot ensembles},}\ }\href {\doibase 10.1103/PhysRevB.86.115308}
  {\bibfield  {journal} {\bibinfo  {journal} {Phys. Rev. B}\ }\textbf {\bibinfo
  {volume} {86}},\ \bibinfo {pages} {115308} (\bibinfo {year}
  {2012})}\BibitemShut {NoStop}%
\bibitem [{\citenamefont {Sinitsyn}\ and\ \citenamefont
  {Pershin}(2016)}]{SinitsynReview}%
  \BibitemOpen
  \bibfield  {author} {\bibinfo {author} {\bibfnamefont {N.~A.}\ \bibnamefont
  {Sinitsyn}}\ and\ \bibinfo {author} {\bibfnamefont {Y.~V.}\ \bibnamefont
  {Pershin}},\ }\bibfield  {title} {\enquote {\bibinfo {title} {The theory of
  spin noise spectroscopy: a review},}\ }\href@noop {} {\bibfield  {journal}
  {\bibinfo  {journal} {Reports on Progress in Physics}\ }\textbf {\bibinfo
  {volume} {79}},\ \bibinfo {pages} {106501} (\bibinfo {year}
  {2016})}\BibitemShut {NoStop}%
\bibitem [{\citenamefont {Poltavtsev}\ \emph {et~al.}(2014)\citenamefont
  {Poltavtsev}, \citenamefont {Ryzhov}, \citenamefont {Glazov}, \citenamefont
  {Kozlov}, \citenamefont {Zapasskii}, \citenamefont {Kavokin}, \citenamefont
  {Lagoudakis}, \citenamefont {Smirnov},\ and\ \citenamefont
  {Ivchenko}}]{PhysRevB.89.081304}%
  \BibitemOpen
  \bibfield  {author} {\bibinfo {author} {\bibfnamefont {S.~V.}\ \bibnamefont
  {Poltavtsev}}, \bibinfo {author} {\bibfnamefont {I.~I.}\ \bibnamefont
  {Ryzhov}}, \bibinfo {author} {\bibfnamefont {M.~M.}\ \bibnamefont {Glazov}},
  \bibinfo {author} {\bibfnamefont {G.~G.}\ \bibnamefont {Kozlov}}, \bibinfo
  {author} {\bibfnamefont {V.~S.}\ \bibnamefont {Zapasskii}}, \bibinfo {author}
  {\bibfnamefont {A.~V.}\ \bibnamefont {Kavokin}}, \bibinfo {author}
  {\bibfnamefont {P.~G.}\ \bibnamefont {Lagoudakis}}, \bibinfo {author}
  {\bibfnamefont {D.~S.}\ \bibnamefont {Smirnov}}, \ and\ \bibinfo {author}
  {\bibfnamefont {E.~L.}\ \bibnamefont {Ivchenko}},\ }\bibfield  {title}
  {\enquote {\bibinfo {title} {Spin noise spectroscopy of a single quantum well
  microcavity},}\ }\href {\doibase 10.1103/PhysRevB.89.081304} {\bibfield
  {journal} {\bibinfo  {journal} {Phys. Rev. B}\ }\textbf {\bibinfo {volume}
  {89}},\ \bibinfo {pages} {081304} (\bibinfo {year} {2014})}\BibitemShut
  {NoStop}%
\bibitem [{\citenamefont {Glasenapp}\ \emph {et~al.}(2013)\citenamefont
  {Glasenapp}, \citenamefont {Greilich}, \citenamefont {Ryzhov}, \citenamefont
  {Zapasskii}, \citenamefont {Yakovlev}, \citenamefont {Kozlov},\ and\
  \citenamefont {Bayer}}]{GlasenappPRB13}%
  \BibitemOpen
  \bibfield  {author} {\bibinfo {author} {\bibfnamefont {P.}~\bibnamefont
  {Glasenapp}}, \bibinfo {author} {\bibfnamefont {A.}~\bibnamefont {Greilich}},
  \bibinfo {author} {\bibfnamefont {I.~I.}\ \bibnamefont {Ryzhov}}, \bibinfo
  {author} {\bibfnamefont {V.~S.}\ \bibnamefont {Zapasskii}}, \bibinfo {author}
  {\bibfnamefont {D.~R.}\ \bibnamefont {Yakovlev}}, \bibinfo {author}
  {\bibfnamefont {G.~G.}\ \bibnamefont {Kozlov}}, \ and\ \bibinfo {author}
  {\bibfnamefont {M.}~\bibnamefont {Bayer}},\ }\bibfield  {title} {\enquote
  {\bibinfo {title} {Resources of polarimetric sensitivity in spin noise
  spectroscopy},}\ }\href {\doibase 10.1103/PhysRevB.88.165314} {\bibfield
  {journal} {\bibinfo  {journal} {Phys. Rev. B}\ }\textbf {\bibinfo {volume}
  {88}},\ \bibinfo {pages} {165314} (\bibinfo {year} {2013})}\BibitemShut
  {NoStop}%
\bibitem [{\citenamefont {Greilich}\ \emph {et~al.}(2006)\citenamefont
  {Greilich}, \citenamefont {Schwab}, \citenamefont {Berstermann},
  \citenamefont {Auer}, \citenamefont {Oulton}, \citenamefont {Yakovlev},
  \citenamefont {Stavarache}, \citenamefont {Reuter}, \citenamefont {Wieck},\
  and\ \citenamefont {Bayer}}]{GreilichPRB73}%
  \BibitemOpen
  \bibfield  {author} {\bibinfo {author} {\bibfnamefont {A.}~\bibnamefont
  {Greilich}}, \bibinfo {author} {\bibfnamefont {M.}~\bibnamefont {Schwab}},
  \bibinfo {author} {\bibfnamefont {T.}~\bibnamefont {Berstermann}}, \bibinfo
  {author} {\bibfnamefont {T.}~\bibnamefont {Auer}}, \bibinfo {author}
  {\bibfnamefont {R.}~\bibnamefont {Oulton}}, \bibinfo {author} {\bibfnamefont
  {D.~R.}\ \bibnamefont {Yakovlev}},
  \bibinfo {author} {\bibfnamefont {M.}~\bibnamefont {Bayer}},
  \bibinfo {author} {\bibfnamefont {V.}~\bibnamefont {Stavarache}},
  \bibinfo {author} {\bibfnamefont {D.}~\bibnamefont {Reuter}},
  \ and\ \bibinfo {author} {\bibfnamefont {A.}~\bibnamefont {Wieck}},
  \ }\bibfield  {title} {\enquote {\bibinfo {title}
  {Tailored quantum dots for entangled photon pair creation},}\ }\href@noop {}
  {\bibfield  {journal} {\bibinfo  {journal} {Phys. Rev. B}\ }\textbf {\bibinfo
  {volume} {73}},\ \bibinfo {pages} {045323} (\bibinfo {year}
  {2006})}\BibitemShut {NoStop}%
\bibitem [{\citenamefont {Warburton}\ \emph {et~al.}(1997)\citenamefont
  {Warburton}, \citenamefont {D\"urr}, \citenamefont {Karrai}, \citenamefont
  {Kotthaus}, \citenamefont {Medeiros-Ribeiro},\ and\ \citenamefont
  {Petroff}}]{PhysRevLett.79.5282}%
  \BibitemOpen
  \bibfield  {author} {\bibinfo {author} {\bibfnamefont {R.~J.}\ \bibnamefont
  {Warburton}}, \bibinfo {author} {\bibfnamefont {C.~S.}\ \bibnamefont
  {D\"urr}}, \bibinfo {author} {\bibfnamefont {K.}~\bibnamefont {Karrai}},
  \bibinfo {author} {\bibfnamefont {J.~P.}\ \bibnamefont {Kotthaus}}, \bibinfo
  {author} {\bibfnamefont {G.}~\bibnamefont {Medeiros-Ribeiro}}, \ and\
  \bibinfo {author} {\bibfnamefont {P.~M.}\ \bibnamefont {Petroff}},\
  }\bibfield  {title} {\enquote {\bibinfo {title} {Charged excitons in
  self-assembled semiconductor quantum dots},}\ }\href {\doibase
  10.1103/PhysRevLett.79.5282} {\bibfield  {journal} {\bibinfo  {journal}
  {Phys. Rev. Lett.}\ }\textbf {\bibinfo {volume} {79}},\ \bibinfo {pages}
  {5282--5285} (\bibinfo {year} {1997})}\BibitemShut {NoStop}%
\bibitem [{\citenamefont {Zapasskii}\ \emph {et~al.}(2013)\citenamefont
  {Zapasskii}, \citenamefont {Greilich}, \citenamefont {Crooker}, \citenamefont
  {Li}, \citenamefont {Kozlov}, \citenamefont {Yakovlev}, \citenamefont
  {Reuter}, \citenamefont {Wieck},\ and\ \citenamefont
  {Bayer}}]{PhysRevLett.110.176601}%
  \BibitemOpen
  \bibfield  {author} {\bibinfo {author} {\bibfnamefont {V.~S.}\ \bibnamefont
  {Zapasskii}}, \bibinfo {author} {\bibfnamefont {A.}~\bibnamefont {Greilich}},
  \bibinfo {author} {\bibfnamefont {S.~A.}\ \bibnamefont {Crooker}}, \bibinfo
  {author} {\bibfnamefont {Yan}\ \bibnamefont {Li}}, \bibinfo {author}
  {\bibfnamefont {G.~G.}\ \bibnamefont {Kozlov}}, \bibinfo {author}
  {\bibfnamefont {D.~R.}\ \bibnamefont {Yakovlev}}, \bibinfo {author}
  {\bibfnamefont {D.}~\bibnamefont {Reuter}}, \bibinfo {author} {\bibfnamefont
  {A.~D.}\ \bibnamefont {Wieck}}, \ and\ \bibinfo {author} {\bibfnamefont
  {M.}~\bibnamefont {Bayer}},\ }\bibfield  {title} {\enquote {\bibinfo {title}
  {Optical spectroscopy of spin noise},}\ }\href {\doibase
  10.1103/PhysRevLett.110.176601} {\bibfield  {journal} {\bibinfo  {journal}
  {Phys. Rev. Lett.}\ }\textbf {\bibinfo {volume} {110}},\ \bibinfo {pages}
  {176601} (\bibinfo {year} {2013})}\BibitemShut {NoStop}%
\bibitem [{\citenamefont {Moody}\ \emph {et~al.}(2014)\citenamefont {Moody},
  \citenamefont {Feng}, \citenamefont {McDonald}, \citenamefont {Mirin},\ and\
  \citenamefont {Silverman}}]{PhysRevB.90.205306}%
  \BibitemOpen
  \bibfield  {author} {\bibinfo {author} {\bibfnamefont {G.}~\bibnamefont
  {Moody}}, \bibinfo {author} {\bibfnamefont {M.}~\bibnamefont {Feng}},
  \bibinfo {author} {\bibfnamefont {C.}~\bibnamefont {McDonald}}, \bibinfo
  {author} {\bibfnamefont {R.~P.}\ \bibnamefont {Mirin}}, \ and\ \bibinfo
  {author} {\bibfnamefont {K.~L.}\ \bibnamefont {Silverman}},\ }\bibfield
  {title} {\enquote {\bibinfo {title} {Homogeneous linewidth narrowing of the
  charged exciton via nuclear spin screening in an $\rm{InAs/GaAs}$ quantum dot
  ensemble},}\ }\href {\doibase 10.1103/PhysRevB.90.205306} {\bibfield
  {journal} {\bibinfo  {journal} {Phys. Rev. B}\ }\textbf {\bibinfo {volume}
  {90}},\ \bibinfo {pages} {205306} (\bibinfo {year} {2014})}\BibitemShut
  {NoStop}%
\bibitem [{\citenamefont {Yang}\ \emph {et~al.}(2014)\citenamefont {Yang},
  \citenamefont {Glasenapp}, \citenamefont {Greilich}, \citenamefont {Reuter},
  \citenamefont {Wieck}, \citenamefont {Yakovlev}, \citenamefont {Bayer},\ and\
  \citenamefont {Crooker}}]{NatComm14}%
  \BibitemOpen
  \bibfield  {author} {\bibinfo {author} {\bibfnamefont {Luyi}\ \bibnamefont
  {Yang}}, \bibinfo {author} {\bibfnamefont {P.}~\bibnamefont {Glasenapp}},
  \bibinfo {author} {\bibfnamefont {A.}~\bibnamefont {Greilich}}, \bibinfo
  {author} {\bibfnamefont {D.}~\bibnamefont {Reuter}}, \bibinfo {author}
  {\bibfnamefont {A.~D.}\ \bibnamefont {Wieck}}, \bibinfo {author}
  {\bibfnamefont {D.~R.}\ \bibnamefont {Yakovlev}}, \bibinfo {author}
  {\bibfnamefont {M.}~\bibnamefont {Bayer}}, \ and\ \bibinfo {author}
  {\bibfnamefont {S.~A.}\ \bibnamefont {Crooker}},\ }\bibfield  {title}
  {\enquote {\bibinfo {title} {Two-colour spin noise spectroscopy and
  fluctuation correlations reveal homogeneous linewidths within quantum-dot
  ensembles},}\ }\href@noop {} {\bibfield  {journal} {\bibinfo  {journal}
  {Nature Communications}\ }\textbf {\bibinfo {volume} {5}},\ \bibinfo {pages}
  {4949} (\bibinfo {year} {2014})}\BibitemShut {NoStop}%
\bibitem [{Note1()}]{Note1}%
  \BibitemOpen
  \bibinfo {note} {Assuming that the pump beam is completely absorbed, and all
  the photoexcited carriers are captured in the topmost QD layer we estimate
  $G\approx P/(\hbar \omega n)$, where $\omega $ is the frequency of the pump
  light. For the $p$-doped sample the generation and recombination rates become
  comparable at $P=0.5$~W/cm$^2$. This value corresponds to $R\approx
  G=0.2$~ns$^{-1}$, which supports our assumption that the recombination and
  generation processes at high pump power are faster than the localized carrier
  spin relaxation. For the $n$-doped sample, however, the estimate gives
  comparable values of $R^{-1}$ and $\tau _s^{exc}$. Our aim is to describe the
  spin noise spectra rather than the generation-recombination dynamics, so we
  use Eq.~\protect \textup {\hbox {\mathsurround \z@ \protect \normalfont
  (\ignorespaces \ref {spectrum}\unskip \@@italiccorr )}} to fit both the
  electron and hole spin noise spectra.}\BibitemShut {Stop}%
\bibitem [{\citenamefont {Kurtze}\ \emph {et~al.}(2009)\citenamefont {Kurtze},
  \citenamefont {Seebeck}, \citenamefont {Gartner}, \citenamefont {Yakovlev},
  \citenamefont {Reuter}, \citenamefont {Wieck}, \citenamefont {Bayer},\ and\
  \citenamefont {Jahnke}}]{Kurtze2009}%
  \BibitemOpen
  \bibfield  {author} {\bibinfo {author} {\bibfnamefont {H.}~\bibnamefont
  {Kurtze}}, \bibinfo {author} {\bibfnamefont {J.}~\bibnamefont {Seebeck}},
  \bibinfo {author} {\bibfnamefont {P.}~\bibnamefont {Gartner}}, \bibinfo
  {author} {\bibfnamefont {D.~R.}\ \bibnamefont {Yakovlev}}, \bibinfo {author}
  {\bibfnamefont {D.}~\bibnamefont {Reuter}}, \bibinfo {author} {\bibfnamefont
  {A.~D.}\ \bibnamefont {Wieck}}, \bibinfo {author} {\bibfnamefont
  {M.}~\bibnamefont {Bayer}}, \ and\ \bibinfo {author} {\bibfnamefont
  {F.}~\bibnamefont {Jahnke}},\ }\bibfield  {title} {\enquote {\bibinfo {title}
  {Carrier relaxation dynamics in self-assembled semiconductor quantum dots},}\
  }\href@noop {} {\bibfield  {journal} {\bibinfo  {journal} {Phys. Rev. B}\
  }\textbf {\bibinfo {volume} {80}},\ \bibinfo {pages} {235319} (\bibinfo
  {year} {2009})}\BibitemShut {NoStop}%
\end{thebibliography}

%

\end{document}